\documentclass{article}
\pdfoutput=1
\usepackage{spconf,amsmath,graphicx,hyperref}
\usepackage{amssymb}
\usepackage{microtype} 

\title{OTCR: Optimal Transmission, Compression and Representation for Multimodal Information Extraction}
\name{A. Yang Li$^{1}$ \quad B. Yajiao Wang$^{1}$ \quad C. Wenhao Hu$^{2}$ \quad D. Zhixiong Zhang$^{1*}$ \quad E. Mengting Zhang$^{1}$ }

\address{
    $^{1}$ University of Chinese Academy of Sciences, Department of Information Resources Management \\
    $^{2}$ University of Electronic Science and Technology of China
}

\begin{document}
\ninept
\maketitle
\begin{abstract} 
Multimodal Information Extraction (MIE) requires fusing text and visual cues from visually rich documents. While recent methods have advanced multimodal representation learning, most implicitly assume modality equivalence or treat modalities in a largely uniform manner, still relying on generic fusion paradigms. This often results in indiscriminate incorporation of multimodal signals and insufficient control over task-irrelevant redundancy, which may in turn limit generalization. We revisit MIE from a task-centric view: text should dominate, vision should selectively support. We present OTCR, a two-stage framework. First, Cross-modal Optimal Transport (OT) yields sparse, probabilistic alignments between text tokens and visual patches, with a context-aware gate controlling visual injection. Second, a Variational Information Bottleneck (VIB) compresses fused features, filtering task-irrelevant noise to produce compact, task-adaptive representations. On FUNSD, OTCR achieves 91.95\% SER and 91.13\% RE, while on XFUND (ZH), it reaches 91.09\% SER and 94.20\% RE, demonstrating competitive performance across datasets. Feature-level analyses further confirm reduced modality redundancy and strengthened task signals. This work offers an interpretable, information-theoretic paradigm for controllable multimodal fusion in document AI.
\end{abstract}

\section{Introduction}
\label{sec:intro}
Multimodal Information Extraction \cite{cui2021document} aims to extract structured knowledge from complex data containing both textual and visual signals. Unlike traditional text-based IE, MIE requires not only understanding natural language but also incorporating modalities such as layout structures, visual symbols, and image elements to mitigate ambiguity and information loss when relying solely on text. This task holds practical value in large-scale scenarios such as invoices \cite{liu2019graph}, receipts \cite{park2019cord}, reports, and form-like documents \cite{jaume2019funsd}, while also serving as a foundation for broader multimodal tasks like document understanding. At the same time, it concerns fundamental methodologies such as cross-modal representation, alignment, and information compression, making it broadly meaningful for advancing multimodal understanding.

In recent years, MIE has achieved substantial progress, with the LayoutLM family \cite{xu2020layoutlm,huang2022layoutlmv3} of pre-trained models and graph-based as well as large language model LLM–based approaches continually pushing SOTA results on multiple benchmarks \cite{cui2021document,xu2022xfund}. Nevertheless, a long-overlooked tension persists: From the perspective of task essence, the data carriers of MIE are mostly document-rich scenarios, where downstream tasks primarily operate on text and its semantic relations. In contrast, layout structures, symbols, or other visual cues mainly serve as anchors for positioning and auxiliary discrimination\cite{bai2022wukong}. Therefore, the modality interaction in MIE should be more appropriately understood as a “text-dominant, vision-supplementary” relationship. However, existing methods often fail to explicitly model this asymmetry, leading to insufficient discrimination and constraint over the relative value of different modalities during the fusion stage. This gives rise to two non-trivial consequences: (i) At the local level, models may develop spurious dependencies on superficial features such as layout, color, or texture, resulting in the feature space being contaminated by unnecessary modal information; (ii) At the global level, the fused representations are often contaminated with task-irrelevant redundancy and noise, inflating the representational space and degrading generalization.

To address this, the core challenge is how to inject visual information into textual features in a controllable and interpretable manner, rather than through simple concatenation or brute-force attention, as well as how to retain only complementary information that benefits the task after fusion.   These challenges are jointly governed by selective filtering, transmission, and compression.  Accordingly, we propose a two-stage cross-modal representation framework, OTCR.   In the first stage, a Cross-modal Information Optimal Transport mechanism captures complementary alignments between text tokens and visual patches across multiple subspaces, injecting visual signals into text in a probabilistically weighted manner while dynamically modulating the degree of visual supplementation.   In the second stage, a variational information bottleneck constrains fused representations, filtering redundancy and noise introduced during alignment and preserving only task-relevant cross-modal signals, thereby producing compact, task-adaptive representations.

The main contributions of this paper are as follows:
\begin{itemize}
    \item We propose a Cross-modal Information Optimal Transport mechanism that establishes minimum-cost alignment path between textual tokens and visual patches, with a context-aware gating design for controllable, semantically relevant visual supplementation.  To the best of our knowledge, this is the first work introducing OT theory into MIE to explicitly address alignment under modality contribution asymmetry.
    \item We design a Variational Information Bottleneck training strategy grounded in information theory, which formalizes the ubiquitous redundancy and noise in cross-modal fusion as an information filtering task. By constraining and compressing the fused representations, our approach effectively preserves the information most valuable for downstream tasks.
    \item Extensive experiments demonstrate the robustness and performance of OTCR compared to current models. We provide detailed metrics and results, underscoring the effectiveness of our approach. Our code will be available at \url{https://github.com/aircraft-young/OTCR}.
\end{itemize}

\section{RELATED WORK}
\label{sec:format}

Existing MIE methods can roughly be divided into four lines: Early \textbf{grid-based approaches} \cite{denk2019bertgrid,dang2021end} attempted to embed textual semantics directly into a 2D layout space, preserving both content and structure at the input level. LiuGraph \cite{liu2019graph} offered another perspective by modeling documents as \textbf{node–edge graphs}, shifting research attention toward more effective graph designs \cite{zhang2022multimodal,tang2021matchvie}. Subsequently, \textbf{large-scale pre-trained models} such as the LayoutLM series \cite{xu2020layoutlm,huang2022layoutlmv3}, together with more recent \textbf{MLLM methods} \cite{he2023icl,ye2023mplug}, have unified text, layout, and vision within a single framework, further advancing cross-task generalization.

Despite methodological differences, existing studies commonly aim to learn a robust multimodal representation to support downstream tasks. ViBERTgrid \cite{lin2021vibertgrid} integrates BERTgrid \cite{denk2019bertgrid} with intermediate CNN layers to enable cross-modal interaction; GraphRevisedIE \cite{cao2023graphrevisedie} employs graph revision techniques to combine multimodal embeddings with global contextual information; DocFormer \cite{appalaraju2021docformer} leverages carefully designed multi-task unsupervised pre-training to enhance cross-modal alignment; and DocReL \cite{li2022relational} introduces relation consistency modeling to generate more effective relational representations.

However, most existing studies tend to directly adopt alignment and fusion paradigms from general multimodal tasks, with limited consideration of the asymmetric contributions of different modalities in MIE. PICK \cite{yu2021pick} and FormNet \cite{lee2022formnet} have recognized that indiscriminate injection of visual information may impair semantic representations, and attempted to mitigate this through graph sparsification or contrastive learning to suppress noise and highlight key relations. Yet these efforts still fall short of explicitly modeling modality asymmetry and they lack a systematic mechanism for compressing redundant information after fusion. Therefore, there is a pressing need for a framework that emphasizes selective transmission and information compression to address the limitations of existing approaches in modeling modality value.

\section{METHODOLOGY}
\label{sec:pagestyle}

\subsection{Problem Formulation}

Given a document image $I$, we obtain a sequence of tokens $\{t_i\}_{i=1}^N$ together with their bounding boxes $\{b_i\}_{i=1}^N$, where each $b_i=(x^0_i,y^0_i,x^1_i,y^1_i)$ denotes the coordinates of token $t_i$. The goal of MIE is to assign entity labels to tokens $\mathcal{C}$ and to predict the relation type between a pair of entities. Therefore, we formulate this task as predicting probability distributions over task-specific label spaces for either tokens or entity pairs.

 In this paper, we aim to address uncontrolled information transmission during cross-modal alignment and fusion, as well as redundancy-induced noise in the fused representation. To this end, a compact representation $\mathcal{Z}$ is learned to preserve task-relevant semantics from across modalities while compressing redundancy. $\hat{y}_i$ serves as the final prediction, including each token’s entity label $t_i$ and relations between entity pairs, this process can be formalized as:
\begin{equation}
    \mathcal{Z} = g(\mathcal{T}, \mathcal{V}), \quad \hat{y}_i = f(\mathcal{Z}_i;\Theta),
\end{equation}
where $g(\cdot)$ denotes the cross-modal transmission, alignment, and compression, and $f(\cdot)$ is a prediction network parameterized by $\Theta$.

\subsection{Overview}

\begin{figure*}
    \centering
    \includegraphics[width=1\textwidth]{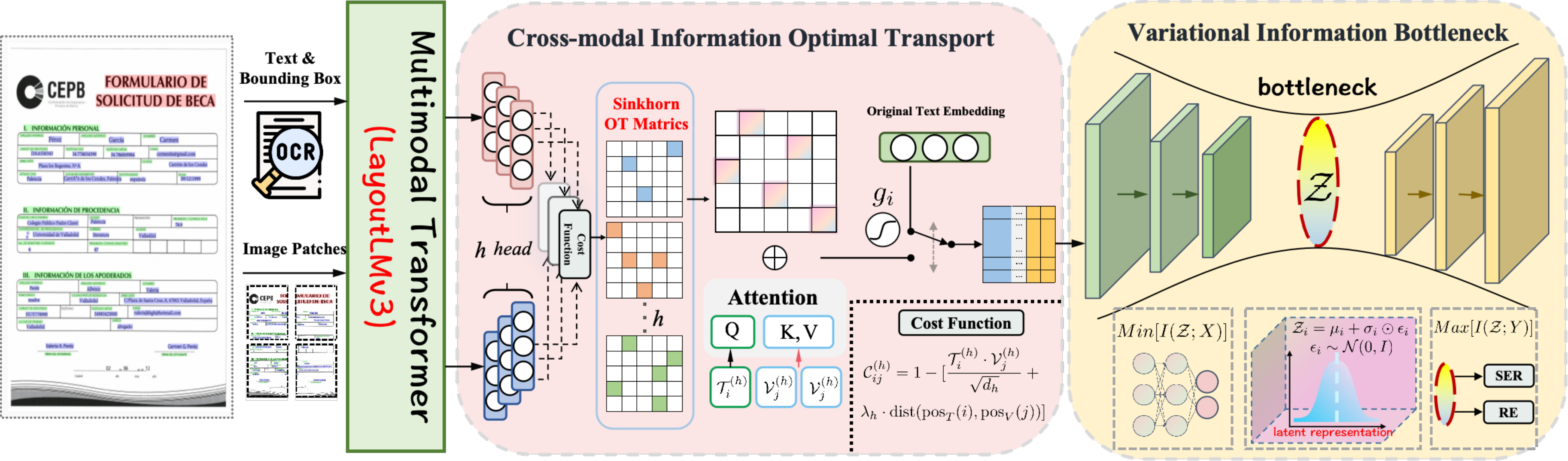}
    \caption{The overall framework of OTCR, which integrates Cross-modal Optimal Transport for controllable visual-to-text injection and a Variational Information Bottleneck for redundancy filtering and task-relevant representation learning.}
    \label{fig:framework}
\end{figure*}

Our framework is illustrated in Figure.~\ref{fig:framework}. Section 3.3 introduces a cross-modal alignment mechanism based on optimal transport, which, under the premise of text dominance, guides visual patches to be selectively injected into textual representations along optimal transport paths, while controlling the proportion of visual supplementation to avoid semantic dilution and noise interference. Section 3.4 further proposes a  variational information bottleneck training strategy to compress and filter the fused representation, retaining only task-relevant complementary semantics and further performing information-theoretic, task-oriented multimodal feature selection.

\subsection{Cross-modal Information Optimal Transport}

To obtain a strong initial representation, we first employ an OCR system to extract textual tokens along with their 2D bounding boxes, and feed them into LayoutLMv3 \cite{huang2022layoutlmv3} for encoding, thereby obtaining the joint text–layout representations $\mathcal{T} \in \mathbb{R}^{N \times d}$ and the visual representations $\mathcal{V} \in \mathbb{R}^{M \times d}$. To capture diverse cross-modal matching patterns across different subspaces, we design a multi-view cross-modal soft alignment mechanism, which is formalized as a multi-head optimal transport coupling problem between two discrete distributions. For the $h$-th head, we apply learnable projections $\mathcal{W}_h^T$ and $\mathcal{W}_h^V$ to map the representations into a specific subspace:
\begin{equation}
\mathcal{T}^{(h)} = \mathcal{T}\mathcal{W}_h^T, 
\quad 
\mathcal{V}^{(h)} = \mathcal{V}\mathcal{W}_h^V,
\end{equation}

Next, each head $h$ defines a parametric cost function $c_h(t_i, v_j)$ and constructs a head-specific cost matrix $\mathcal{C}^{(h)}$:
\begin{equation}
\mathcal{C}^{(h)}_{ij} = -\frac{\mathcal{T}^{(h)}_i \cdot \mathcal{V}^{(h)}_j}{\sqrt{d_h}}
+ \lambda_h \cdot \mathrm{dist}\!\left(\mathrm{pos}_T(i), \mathrm{pos}_V(j)\right),
\end{equation}
where the first term measures semantic similarity, and the second term introduces a learnable spatial bias $\lambda_h$ to capture layout-aware proximity. $\mathrm{pos}_T(i)$ and $\mathrm{pos}_V(j)$ denote the normalized 2D coordinates of the $i$-th text token and the $j$-th visual patch, respectively. Thus, each head encodes a distinct alignment principle. To find the optimal transmission path of text and visual information from different perspectives, we adopt the Sinkhorn algorithm within entropy-regularized optimal transport. By alternately scaling the rows and columns, Sinkhorn \cite{cuturi2013sinkhorn} avoids collapse into a few matches and yields a globally consistent and mass-conserving soft alignment. The resulting transport plan is a doubly stochastic matrix:
\begin{equation} \pi^{(h)} = \mathrm{diag}({a}) \, \exp\!\left(-\frac{\mathcal{C}^{(h)}}{\tau}\right) \, \mathrm{diag}({b}), \end{equation}
where $\pi^{(h)}_{ij}$ denotes the probability mass assigned from visual token $j$ to text token $i$ under the $h$-th head. Finally, we aggregate the alignments across all heads by equal-weighted averaging:
\begin{equation}
\mathcal{P} = \frac{1}{H} \sum_{h=1}^H \pi^{(h)}.
\end{equation}

After obtaining the cross-modal alignment matrix $\mathcal{P}$, we combine two complementary sources of information: a global attention path and a local OT aggregation path:
\begin{equation}
\mathcal{F}_{att} = {MultiHeadAttn}(\mathcal{T}, \mathcal{V}, \mathcal{V}),\quad
\end{equation}
\begin{equation}
\mathcal{F}_{ot}(i) = \sum_j \mathcal{T}_i \cdot \mathcal{P}_{ij} \cdot \mathcal{V}_j,
\end{equation}
\begin{equation}
\mathcal{F}_{fusion} = \mathcal{F}_{att} + \mathcal{F}_{ot}.
\end{equation}

Next, to ensure that visual information can be selectively injected with matching text information, we introduce a context-aware gating mechanism that dynamically controls the contribution of visual features. Formally, for each text token $i$, the gate value $g_i$ is computed based on its original text embedding, the fused feature, and the confidence of the alignment distribution:
\begin{equation}
g_i = \sigma\!\left(\mathcal{W}_g \,[\mathcal{T}_i ; \mathcal{F}_{fusion}(i) ; \text{conf}_i]\right),
\end{equation}
where $\sigma$ is the sigmoid function, and $\text{conf}_i = -\sum_j \mathcal{P}_{ij} \log \mathcal{P}_{ij}$ represents the entropy of the alignment distribution, indicating the uncertainty of how text token $i$ matches with visual tokens. A higher entropy means lower confidence and thus a smaller gate value. The final gated representation is obtained as:
\begin{equation}
\mathcal{T}'_i = g_i \cdot \mathcal{F}_{fusion}(i) + (1 - g_i) \cdot \mathcal{T}_i.
\end{equation}
This allows the model to adaptively balance text semantic dominance and visual enhancement: when the alignment is confident, the model incorporates more visual evidence; when the alignment is uncertain, the model preserves the original text semantics.

\subsection{Variational Information Bottleneck Training Strategy for Redundancy Suppression}

Token–patch alignment alone cannot ensure that the fused representation is compact and redundancy-free for the downstream task. In other words, even though the injection of visual information is controlled, the overall representation may still retain redundant or noisy signals, which undermines generalization. To address this, we learn an intermediate representation $\mathcal{Z}$ that is sufficient and minimal: it preserves task-relevant information (maximize $I(\mathcal{Z}; Y)$) while compressing redundancy from the input (minimize $I(\mathcal{Z}; X)$). 

Taking the gated fusion representation $\mathcal{T}'_i$ of each token as the input to the bottleneck, a variational encoder parameterizes a Gaussian distribution for each token:
\begin{equation}
\mu_i = W_{\mu}\mathcal{T}'_i + b_{\mu}, 
\quad
\log \sigma_i^2 = W_{\sigma}\mathcal{T}'_i + b_{\sigma},
\end{equation}
Using the reparameterization trick, we draw a latent representation:
\begin{equation}
\mathcal{Z}_i = \mu_i + \sigma_i \odot \epsilon_i, 
\quad \epsilon_i \sim \mathcal{N}(0, I).
\end{equation}
The latent $\mathcal{Z}_i$ serves as the intermediate representation for prediction. To enforce the principle of sufficiency and minimality, we adopt the variational information bottleneck objective: the task supervision loss $\mathcal{L}_{task}$ provides a variational lower bound to $I(\mathcal{Z}; Y)$ by maximizing the expected log-likelihood of labels given $\mathcal{Z}$, while a KL regularization term $\mathcal{L}_{VIB}$ serves as an upper bound proxy to $I(\mathcal{Z}; X)$ by pushing the posterior distribution towards an isotropic Gaussian prior to filter redundancy. The task loss is defined as:
\begin{equation}
\begin{aligned}
\mathcal{L}_{task} 
&= -\mathbb{E}_{q(\mathcal{Z}|X)}[\log p_\theta(Y|\mathcal{Z})] \\
&= -\frac{1}{N} \sum_{i=1}^N \sum_{c \in \mathcal{C}} y_{i,c}\,\log \hat{y}_{i,c}.
\end{aligned}
\end{equation}
thus serving as a practical proxy for maximizing $I(\mathcal{Z}; Y)$. The overall optimization objective is:
\begin{equation}
\mathcal{L} = \mathcal{L}_{task} 
+ \beta \cdot \frac{1}{N} \sum_{i=1}^{N} 
\mathrm{KL}\!\left(q(\mathcal{Z}_i|\mathcal{T}'_i)\,\|\,\mathcal{N}(0, I)\right),
\end{equation}
where $\beta$ is a hyperparameter balancing discriminative power and compression. Notably, the closed-form KL solution improves interpretability: irrelevant dimensions collapse to the prior ($\mu \rightarrow 0$ and $\sigma^2 \rightarrow 1$), while relevant ones keep non-zero means and lower variances, showing selective retention of discriminative features.

\section{EXPERIMENTS}
\label{sec:typestyle}
In this section, we evaluate the effectiveness of our model and aim to address the following questions:
\begin{itemize}
    \item \textbf{RQ1:} Does our proposed framework achieve superior performance compared with state-of-the-art methods on MIE tasks?
    \item \textbf{RQ2:} Does OTCR produce more compact and task-relevant multimodal representations by explicitly modeling modality asymmetry and filtering redundancy?
    \item \textbf{RQ3:} Do the different modules within OTCR improve the efficiency of MIE?
\end{itemize}

\subsection{Experiment Settings}
\subsubsection{Datasets and Baselines}
The datasets used in this study for semantic entity recognition (SER) and relation extrction (RE) tasks are FUNSD \cite{jaume2019funsd} and XFUND \cite{xu2022xfund}. FUNSD contains 199 scanned forms with entity and relation annotations, while XFUND extends it to seven languages with token-level labels for semantic entities. We compare OTCR with eleven baseline models, including the grid-based methods: \textbf{MSAU-PAF}~\cite{dang2021end}, and large-scale pre-trained methods: \textbf{LayoutLMv3}~\cite{huang2022layoutlmv3}, \textbf{LayoutXLM}~\cite{xu2021layoutxlm}, \textbf{XYLayoutLM}~\cite{gu2022xylayoutlm}, \textbf{ESP}~\cite{yang2023modeling}, \textbf{RORE}~\cite{zhang2024modeling}, \textbf{DocExtractNet}~\cite{yan2025docextractnet}, \textbf{LiLT}~\cite{wang2022lilt}, \textbf{KVPFormer}~\cite{hu2023question}; and graph-based methods: \textbf{DocGraphLM}~\cite{wang2023docgraphlm}, \textbf{FormNet}~\cite{lee2022formnet}. Following previous methods, we use F1 score as performance metric.

\subsubsection{Implementation Details}
We adopt LayoutLMv3 \cite{huang2022layoutlmv3} as the backbone. Documents are OCR-processed, resized to $224 \times 224$, and encoded into aligned visual–text embeddings. Sequences are padded or truncated to a fixed length with hidden size $768$. Models are trained for $50$ epochs with batch size $12$ and learning rate $4e^{-5}$, using fixed seeds for reproducibility. All experiments ran on 4 NVIDIA A100 (80GB) GPUs.

\subsection{Performance Comparison (RQ1)}
Table~\ref{table:comparison} reports the performance of OTCR against competitive baselines on FUNSD and XFUND(ZH). (i) On FUNSD, OTCR clearly outperforms all methods, reaching the best SER (91.95\%) and RE (91.13\%), which demonstrates its robustness in extracting both entities and relations under complex layouts. (ii) On XFUND(ZH), OTCR achieves a strong RE of 94.20\%, nearly matching KVPFormer (94.27\%), while attaining a solid SER of 91.09\%, close to XYLayoutLM’s 91.76\%, while delivering more balanced improvements across SER and RE. These results indicate that OTCR is not optimized for a single metric but enhances overall representation quality in a more comprehensive manner.
\begin{table}[t]
\caption{Overall comparison results on FUNSD and XFUND(ZH).}
\centering
\label{table:comparison}
\begin{tabular}{lccccc}
\hline
 & \multicolumn{2}{c}{FUNSD} & \multicolumn{2}{c}{XFUND(ZH)} \\
Method & SER & RE & SER & RE \\
\hline
MSAU-PAF \cite{dang2021end}       & 83.00 & 75.00 & --    & --    \\
LayoutLMv3 \cite{huang2022layoutlmv3}      & 90.29 & 86.52 & --    & 93.63 \\
LayoutXLM \cite{xu2021layoutxlm}       & 79.40 & 54.83 & 89.24 & 70.73 \\
XYLayoutLM \cite{gu2022xylayoutlm}     & 83.55 & --    & \textbf{91.76} & 74.45 \\
ESP \cite{yang2023modeling}             & 91.12 & 88.88 & 90.30 & 90.80 \\
RORE \cite{zhang2024modeling}            & 91.84 & 88.46 & --    & --    \\
DocExtractNet \cite{yan2025docextractnet}   & 91.80 & --    & --    & --    \\
LiLT\cite{wang2022lilt}            & --    & --    & 89.36 & 72.97 \\
DocGraphLM \cite{wang2023docgraphlm}      & 88.77 & --    & --    & --    \\
KVPFormer \cite{hu2023question}       & --    & 90.86 & --    & \textbf{94.27} \\
FormNet \cite{lee2022formnet}       & 84.69    & -- & --    & -- \\
\textbf{OTCR(Ours) }           & \textbf{91.95} & \textbf{91.13} & 91.09 & 94.20 \\
\hline
\end{tabular}
\end{table}

\subsection{T-SNE Analysis of Multimodal Representations (RQ2)}
\begin{figure}[t]
    \centering
    \includegraphics[width=0.45\textwidth]{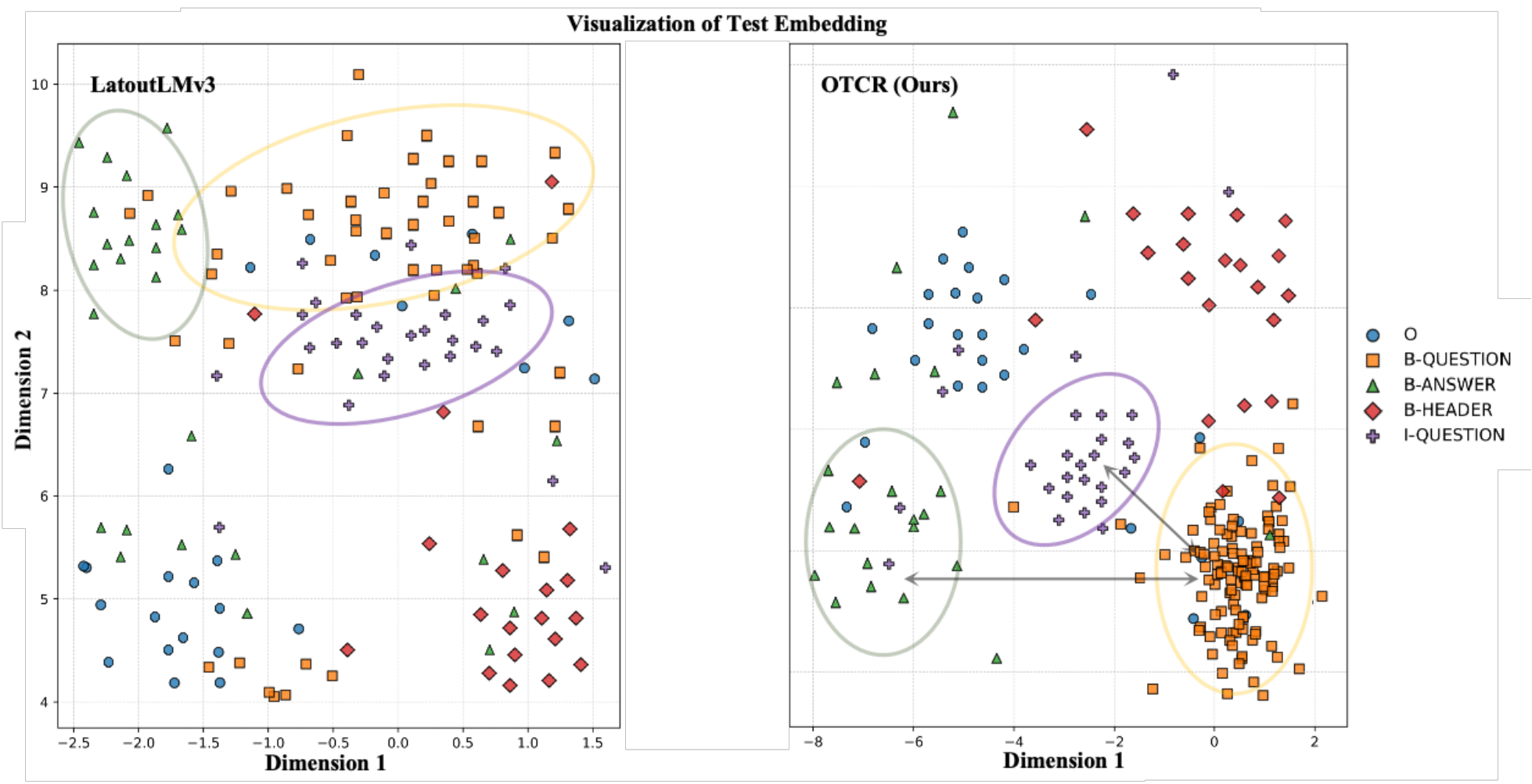}
    \caption{T-SNE visualization of the final-layer hidden embeddings on the FUNSD dataset across different models.}
    \label{fig:cdf}
\end{figure}

To examine whether OTCR learns more compact and task-relevant multimodal representations, we conducted embedding visualization on the FUNSD. Specifically, the T-SNE visualization in Figure~\ref{fig:cdf} shows that LayoutLMv3 embeddings exhibit fragmentation and overlap. In contrast, OTCR produces a much clearer structure: B-Answer and I-Answer form a tightly organized cluster, B-Question remains adjacent but separable, B-Header constitutes an isolated high-margin cluster, and O tokens are effectively pushed to the periphery. These patterns demonstrate that OTCR, through optimal transport alignment and variational information bottleneck, successfully filters redundancy and strengthens task-relevant signals, yielding compact and discriminative multimodal representations.

\subsection{Ablation Study (RQ3)}
\begin{figure}[t]
    \centering
    \includegraphics[width=0.45\textwidth]{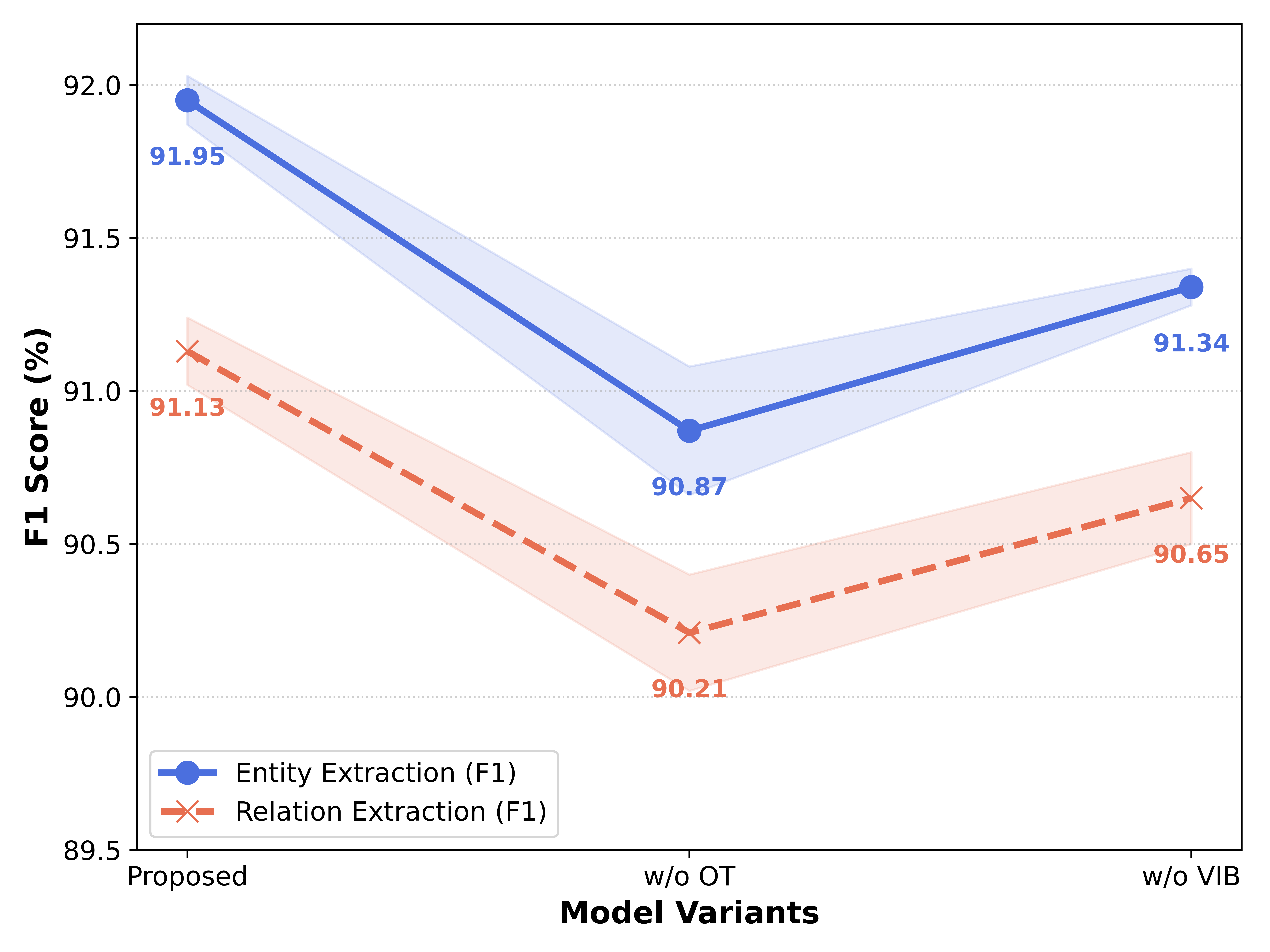}
    \caption{Ablation results on the FUNSD dataset.}
    \label{fig:abl}
\end{figure}

The ablation study on the FUNSD dataset highlights the contributions of each module. As shown in Figure~\ref{fig:abl}, both curves peak with the proposed model, while removing OT or VIB consistently reduces performance. Specifically, removing the cross-modal OT module caused drops of 1.08 and 0.92 points on SER and RE, respectively; removing the variational information bottleneck reduced them by 0.61 and 0.48. These results show that both components are beneficial, but OT has a stronger impact, indicating that explicit OT alignment is the main driver of improvement: without OT, visual information cannot be selectively injected into text representations along optimal paths, causing semantic dilution; in contrast, VIB complements alignment by compressing redundancy and enhancing generalization, thereby yielding stable gains. In addition, shaded regions indicate standard deviation across runs with different seeds (mean ±1 std). Despite fluctuations among variants, the proposed model consistently attains higher mean scores than w/o OT and VIB, confirming that the improvements are stable and reliable.

\section{CONCLUSION}
\label{sec:majhead}
In this work, we proposed OTCR, a framework that explicitly models modality asymmetry via optimal transport alignment and suppresses redundancy through variational information bottleneck.  Experiments on FUNSD and XFUND demonstrate that OTCR achieves highly competitive results in both entity and relation extraction, with ablation studies and visualization further confirming that our “align-then-compress” paradigm produces more compact and task-relevant multimodal representations.

\vfill\pagebreak


\bibliographystyle{IEEEbib}
\bibliography{refs}

\end{document}